\documentclass[lettersize,journal]{IEEEtran}
\usepackage{amsmath}
\usepackage{algorithmic}
\usepackage{algorithm}
\usepackage{array}
\usepackage[caption=false,font=normalsize,labelfont=sf,textfont=sf]{subfig}
\usepackage{textcomp}
\usepackage{stfloats}
\usepackage{url}
\usepackage{verbatim}
\usepackage{graphicx}
\usepackage{cite}
\usepackage[utf8]{inputenc}
\usepackage{amsthm,amssymb,amsfonts,mathtools}
\usepackage{xcolor}
\usepackage{booktabs}
\usepackage{tabularx}
\usepackage{tabulary}
\usepackage{tabularray}
\usepackage{multirow}
\usepackage{stackengine}
\allowdisplaybreaks

\makeatletter
\newcommand{\specialcell}[1]{\ifmeasuring@#1\else\omit$\displaystyle#1$\ignorespaces\fi}
\makeatother

\begin{document}

\title{Assessing the Reliability Benefits of Energy Storage as a Transmission Asset}

\author{David Sehloff,~\IEEEmembership{Member,~IEEE,}
Jonghwan Kwon,~\IEEEmembership{Member,~IEEE,}
Mahdi Mehrtash,~\IEEEmembership{Senior Member,~IEEE,}

Todd Levin,~\IEEEmembership{Member,~IEEE,}
Benjamin F. Hobbs,~\IEEEmembership{Life Fellow,~IEEE}
}

\maketitle

\begin{abstract}
Utilizing energy storage solutions to reduce the need for traditional transmission investments has been recognized by system planners and supported by federal policies in recent years. This work demonstrates the need for detailed reliability assessment for quantitative comparison of the reliability benefits of energy storage and traditional transmission investments. First, a mixed-integer linear programming expansion planning model considering candidate transmission lines and storage technologies is solved to find the least-cost investment decisions. Next, operations under the resulting system configuration are simulated in a probabilistic reliability assessment which accounts for weather-dependent forced outages. The outcome of this work, when applied to TPPs, is to further equalize the consideration of energy storage compared to traditional transmission assets by capturing the value of storage for system reliability.
\end{abstract}


\section{Introduction}
New electricity transmission infrastructure and energy storage installations are recognized as key enablers of meeting decarbonization goals while maintaining the reliability of power systems. Traditionally, the services these assets can provide have been treated as distinct, but recently, energy storage has been considered as a potential source of transmission services. Storage operated as a transmission asset may offer preferable capital costs and construction timelines compared to traditional transmission solutions, while bringing unique advantages for supporting the system and enhancing both reliability and resilience.

The development of regulatory guidelines for this application of energy storage is evolving in the United States. In 2017, the Federal Energy Regulatory Commission (FERC) issued a policy statement allowing dual use of energy storage assets, that is, cost recovery through both the regulated transmission planning practices (TPPs) and participation in competitive markets \cite{federal_energy_regulatory_commission_utilization_2017}. Between 2020 and 2023, FERC issued orders accepting proposals from three independent system operators defining a new class of asset, ''Storage as Transmission-Only Asset'' (SATOA). A storage installation of this type may be selected through TPPs provided that it addresses a specific transmission system need and follows other requirements ensuring it provides only transmission services \cite{federal_energy_regulatory_commission_order_2020, federal_energy_regulatory_commission_order_2023-1, federal_energy_regulatory_commission_order_2023}. As a result, the Midcontinent Independent System Operator considered a SATOA in its TPP and identified it as superior to a transmission line, or \emph{wires}, solution, leading to its selection and plans for permitting and construction \cite{midcontinent_independent_system_operator_2019_2019}. Despite this movement toward consideration of energy storage for transmission services, existing TPPs have barriers and disincentives which make them inadequate for evaluation of energy storage solutions \cite{twitchell_enabling_2022}, one of which is evaluation of the reliability contributions of energy storage compared with traditional wires solutions.

The co-optimization of transmission and energy storage planning has been assessed under various distinct perspectives. One common perspective is the centralized co-planning of transmission and storage to avoid the need for new transmission installations, thereby reducing investment costs \cite{zechun_hu_transmission_2012,macrae_transmission_2014,qi_joint_2015,hedayati_joint_2014,qiu_stochastic_2017}. Variations of this perspective include incorporation of security constraints and high renewable penetration \cite{gan_security_2019}. Konstantelos and Strbac included additional flexible network and non-network solutions in this centralized planning approach and demonstrated the value of these solutions in reducing risk under uncertain evolution of system capacity \cite{konstantelos_valuation_2015}. Go et al. took the perspective of co-planning generation, transmission, and energy storage, showing the benefits of a co-optimization approach to reduce investment costs while meeting renewable portfolio standard constraints \cite{go_assessing_2016}. Dvorkin et al. considered transmission and energy storage co-planning from the perspective of a merchant energy storage owner in a multi-level framework finding maximum energy storage system profit with centralized least-cost transmission expansion decisions \cite{dvorkin_co-planning_2018}. Aside from these approaches considering avoided costs and profits, Zach and Auer compared transmission and energy storage investments with a welfare maximization approach \cite{zach_bulk_2012}. Kwon et al. introduced a framework for optimal sizing and siting of energy storage as a transmission asset in order to satisfy specific needs identified in TPPs \cite{kwon_optimal_2023}. However, previous studies primarily focus on the economic value of transmission and storage investments, lacking consideration of the reliability contributions of these investments. 

Many approaches for assessing the reliability of power systems have been applied to transmission expansion planning. The expected unserved energy (EUE) is commonly used to capture the energy shortfall, for example in \cite{jun_hua_zhao_flexible_2009,hooshmand_combination_2012}, and  metrics such as loss of load expectation (LOLE) and loss of load hours (LOLH) can provide additional insights into the temporal nature of shortfalls \cite{ibanez_comparing_2014}. Similar probabilistic metrics have been introduced, such as loss of load cost (LOLC) \cite{leite_da_silva_reliability_2010}. An alternative to probabilistic reliability metrics is to ensure robustness to a defined set of transmission and generation contingencies through $N-1$ \cite{gu_transmission_2012} or $N-k$\cite{choi_transmission_2007,moreira_reliable_2017} security constraints. The 2022 IEEE Composite Power System Reliability report \cite{bagen_bagen_et_al_composite_2022} highlights the need to develop composite system reliability methods which incorporate storage, common and dependent mode failure models, and weather conditions into composite system reliability analysis. Sundar et al. \cite{sundar_meteorological_2023} applied temperature-dependent forced outage rates to natural gas combined cycle and hydropower plants for resource adequacy assessment, not considering the transmission network. Other studies including \cite{firouzi_reliability_2022,thapa_composite_2023} have considered uncertainty from renewable generation and included representation of the transmission system, but they do not model weather-dependent or correlated generation outages. 

Means for assessing the reliability value of energy storage have been a subject of much recent work, as briefly reviewed in \cite{ayesha_reliability_2023}. Stephen et al. demonstrated that the approach by which storage dispatch is modeled can have a large impact on resource adequacy metrics \cite{stephen_impact_2022}, particularly whether system reliability (as in \cite{evans_minimizing_2019, evans_assessing_2020,bromley-dulfano_reliability_2021,qi_capacity_2024}) or economic performance (as in \cite{dratsas_real-time_2024}) is the assumed dispatch objective. However, these studies focus on resource adequacy with limited or no modeling of the transmission system and do not take into account the concept of SATOA in assessing the reliability values.

This study considers the viewpoint of a system operator co-optimizing transmission and energy storage for least-cost investment and operations and introduces a framework to quantify the reliability contribution that energy storage as a transmission asset provides at a system level and to compare these benefits with those of transmission upgrades.

This study is the continuation of the conference paper \cite{mehrtash_necessity_2023}, which was presented at the Texas Power and Energy Conference (TPEC 2023). The main contribution of  \cite{mehrtash_necessity_2023} was to show the advantage of joint resource and transmission expansion planning in comparison with the resource-only expansion planning and project-by-project screening of expansion plans considering different future scenarios representing uncertainties that exist in emission reduction policies and forecasted demand. This article significantly extends the authors’ previous work  \cite{mehrtash_necessity_2023} by conducting a comparative analysis of reliability outcomes stemming from various investment strategies: focusing solely on transmission, solely on storage, or permitting investments in both. The aim is to gain comprehensive insights into the reliability contributions offered by SATOA and traditional transmission lines.

The contributions of this paper are the following:
\begin{itemize}
    \item An analysis framework that supports the assessment of the economic and reliability value of both SATOA and traditional transmission assets. 
    \item A composite reliability assessment model considering weather-dependent correlated generator outages, pre- and post-contingency operations including storage as a transmission only asset, and the transmission network represented with a DC power flow model
    \item A case study demonstrating this framework for a representative high-renewable, limited-transmission system, including findings highlighting the importance of capturing reliability contributions of transmission assets including storage in transmission planning, demonstrating that SATOA can offer comparable benefits to transmission line investments.
\end{itemize}

The remainder of the paper is organized as follows. Section \ref{methods} presents the methodology for the transmission and energy storage expansion planning and the probabilistic reliability assessment and their coupling, Section \ref{sec:casestudy} presents a case study on a simplified ERCOT-like system, and Section \ref{sec:conclusions} concludes.

\section{Methods}\label{methods}
This section presents an overview of the two-stage analysis framework, as shown in Fig. 1, employed in this study. In the first stage, the Johns Hopkins Stochastic Multi-stage Integrated Network Expansion (JHSMINE) model was utilized for optimal planning of transmission and storage expansion.
In the second stage, the expansion decisions derived from JHSMINE are passed to the Argonne Low-carbon Electricity Analysis Framework (A-LEAF) for a comprehensive probabilistic reliability assessment. 

\begin{figure*}
    \centering
    \includegraphics[width=0.7\textwidth]{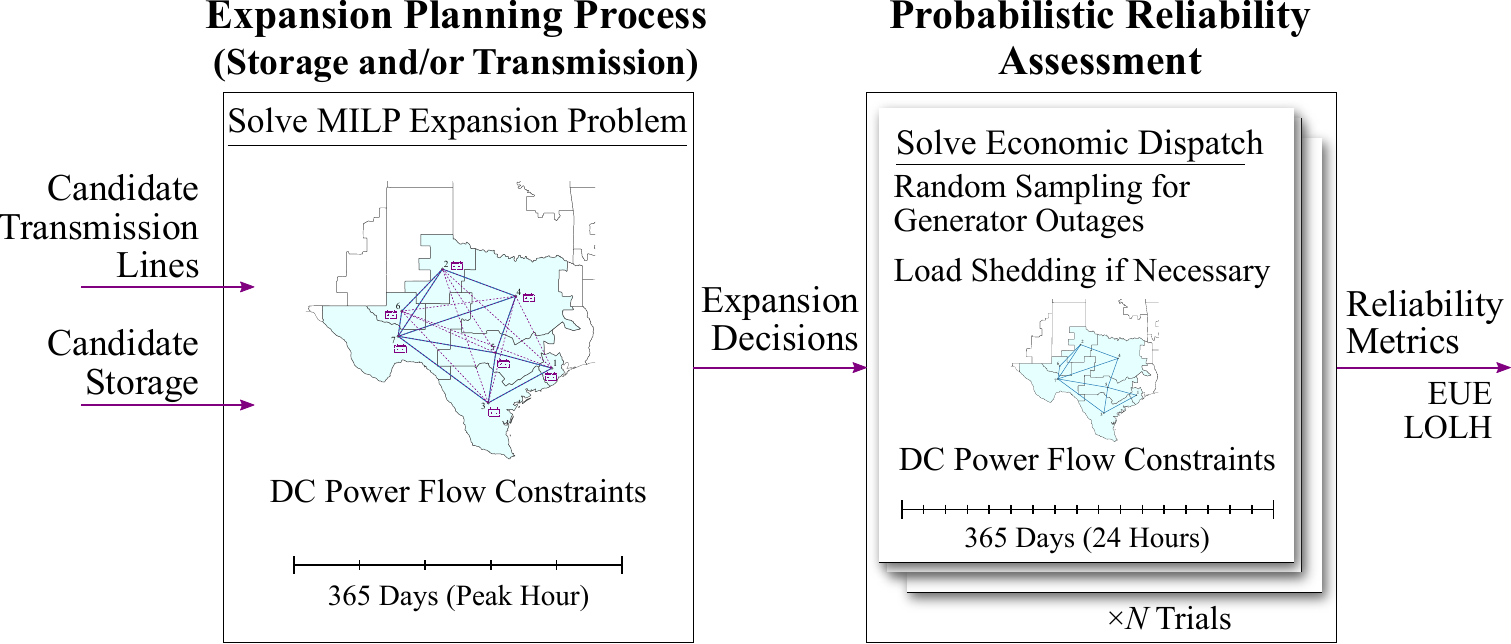}
    \caption{Overview of the analysis framework. The expansion planning process, described in subsection \ref{sec:btep}, selects optimal size and location of storage and transmission assets, and the probabilistic reliability assessment, described in subsection \ref{sec:ra}, simulates operations under random generation outage scenarios and computes reliability metrics.}
    \label{fig:method}
\end{figure*}

\subsection{Transmission and Storage Expansion}\label{sec:btep}
JHSMINE is a cost-minimizing capacity expansion planning tool that considers several significant aspects of the expansion planning models such as unit commitment, linearized power flow, renewable energy policies and energy trading credits, etc., and has been used in several projects \cite{electric_power_research_institute_coordinated_2022, mehrtash_necessity_2023, xu_value_2019}. In this paper, JHSMINE is used to find expansion planning solutions with three different cases. In the first case, only the transmission investment option is active. In the second case, storage investment is the only available option. Lastly, in the third case, both transmission and storage investments are considered as investment options. It is intuitive that conducting joint storage and transmission expansion planning reduces costs relative to the cases that allow either transmission-only or storage-only expansion planning. However, assessing all three of these cases allows us to  evaluate the contribution of each individual technology to system reliability. Although JHSMINE is capable of running a stochastic multistage expansion planning model; here we apply its deterministic expansion planning mode for a single planning horizon.

The objective function of JHSMINE is to minimize the probability-weighted of total system cost as represented by \eqref{JHSMINE:eq1}. The objective function consists of two terms: investment cost and operation cost, and both are realized at the scenario tree node $(s,y)$, where $s$ and $y$ are indices for scenarios and years, respectively. JHSMINE discounts the investment cost back to the beginning of the planning horizon using $D_y$, where $\delta$ is the interest rate. JHSMINE assumes that the operation condition (e.g., load, policies, fuel price, etc.) of operation node will repeat until the next operation simulation interval.  All cash flows for operations are assumed to be end-of-year flows, and the discounting formulas are defined accordingly as $DA_y$.

\vspace{-1.0em}
\small
\begin{gather}
\begin{aligned}
    \omit\rlap{$\min\quad obj=\sum_{\left(s,y\right)}{SP_{s,y}}\left(D_y\cdot invc_{s,y}+DA_y\cdot oprc_{s,y}\right),$}&\\
	\quad &D_y=1/(1+\delta)^{y-y_{origin}},\\
    \quad &DA_y=D_y\cdot P\left(1,Y_y,\delta\right)=D_y\left(\sum_{t=1}^{Y_y}\frac{1}{\left(1+\delta\right)^t}\right)\label{JHSMINE:eq1}
\end{aligned}
\end{gather}
\normalsize

JHSMINE calculates the investment cost of new facilities using \eqref{JHSMINE:eq2}, which includes the expansion costs of facilities that are newly expanded and the salvage revenue for facilities that are newly retired. 

\vspace{-1.0em}
\small
\begin{align}
    invc_{s,y}\!&=\!GEXC_{s,y,k}\!\cdot\! gincexp_{s,y,k}\nonumber\\
    &\hfill-\sum_{\mathclap{\left(s^\prime,y^\prime\right)}}\!{GSAL_{s,y,k}\!\cdot\!gincret_{s,y,k}}+LEXC_{s,y,l}\!\cdot\!lincexp_{s,y,l}\nonumber\\
    &\hfill-\sum_{\mathclap{\left(s^\prime,y^\prime\right)}}\!{LSAL_{s,y,l}\!\cdot\!lincret_{s,y,l}}+EEXC_{s,y,j}\!\cdot\! eincexp_{s,y,j}\nonumber\\
    &\hfill-\sum_{\mathclap{\left(s^\prime,y^\prime\right)}}\!{ESAL_{s,y,j}\!\cdot\!eincret_{s,y,j}}\label{JHSMINE:eq2}
\end{align}
\normalsize

JHSMINE calculates the operation cost using \eqref{JHSMINE:eq3}, which is composed of seven terms: the fixed operation and maintenance (O\&M) cost of generation and storage facilities represented by \eqref{JHSMINE:eq4}; the variable O\&M cost of generation and storage operations modeled by \eqref{JHSMINE:eq5}; the fuel cost of generation and storage operations shown in \eqref{JHSMINE:eq6}; the start-up cost from generation unit commitment represented by \eqref{JHSMINE:eq7}; and finally the value of lost load modeled by \eqref{JHSMINE:eq8}.

\vspace{-1.0em}
\small
\begin{align}
&oprc_{s,y}=fomc_{s,y}+vomc_{s,y}+stuc_{s,y}+voll_{s,y}\label{JHSMINE:eq3}\\
&fomc_{s,y}=\sum_{k}{GFOM_k\cdot GNPL_k\cdot gstat_{s,y,k}}\nonumber\\[-0.5em]
&\specialcell{\hfill+\sum_{j}{EFOM_j\cdot EGCP_j\cdot estat_{s,y,j}}}\label{JHSMINE:eq4}\\
&vomc_{s,y}=\sum_{rp,h}\!RPW_{y,rp}\!\Bigg(\!\sum_{k}\!{GVOM_k\!\cdot\!g o p t_{s,y,rp,h,k}}\nonumber\\[-0.5em]
&\specialcell{\hfill+\sum_{j}{EVOM_j\cdot e d i s_{s,y,rp,h,j}}\Bigg)}\label{JHSMINE:eq5}\\
&fuelc_{s,y}=\sum_{rp,h}RPW_{y,rp}\Bigg[\sum_{f}FC_{s,y,f,rp,h}\nonumber\\[-0.25em]
&\specialcell{\hfill\!\cdot\!\left(\!\sum_{k\in K_f}\!{\!GHR_k\!\cdot\! g o p t_{s,y,rp,h,k}}\!+\!\sum_{\mathclap{j\in J_f}}\!{EHR_j\!\cdot\! e d i s_{s,y,rp,h,j}}\!\right)\!\Bigg]}\label{JHSMINE:eq6}\\
&stuc_{s,y}=\sum_{rp,h}RPW_{y,rp}\nonumber\\[-0.25em]
&\specialcell{\hfill\cdot\left(\sum_{k}{GSUC_k\cdot G N P L_k\cdot g s u p_{s,y,rp,h,k}}\right)}\label{JHSMINE:eq7}\\
&voll_{s,y}=\sum_{rp,h}{RPW_{y,rp}\!\cdot\! V O L L\left(\sum_{i}{nload_{s,y,rp,h,i}}\right)}\label{JHSMINE:eq8}
\end{align}
\normalsize

The feasibility space of JHSMINE is constructed by a set of equality constraints and a set of inequality constraints. Among the equality constraints are nodal power balances in the system, DC power flow representation, energy balances for storage, etc. The set of inequality constraints includes thermal limits of transmission lines, output power limit of generating units, energy storage representations, etc. [29], [30].
The unit commitment constraints in JHSMINE are expanded based on the ``Tight Relaxed Unit Commitment'' (TRUC) Constraints in \cite{kasina_essays_2017}, For classic unit commitment without generation expansion, readers are referred to \cite{baldick_generalized_1995} and \cite{morales-espana_tight_2013}. Under TRUC, unit commitment variables, i.e., operating status, start-up, and shut-down variables, can be relaxed; in the meanwhile, each unit commitment constraint is still physically meaningful. For the power flow representation, a disjunctive version of the DC Optimal Power Flow (DCOPF) is modeled in JHSMINE.
The detailed model formulations of JHSMINE, including the set of constraints, can be found in \cite{mehrtash_jhsmine_2024}.

\subsection{Probabilistic Reliability Assessment}\label{sec:ra}
A-LEAF is an integrated power system simulation framework that includes a suite of generation and transmission capacity expansion, unit commitment, and system reliability assessment models\footnote{\url{https://www.anl.gov/esia/a-leaf}}. In this paper, we employ the probabilistic reliability assessment model in A-LEAF, which conducts Monte Carlo simulations to quantify the risk of unserved energy (i.e., the system’s reliability level) and calculates various reliability metrics.  Furthermore, the reliability assessment model was enhanced to specifically consider the operation of SATOA.

\begin{figure*}
    \centering
    \includegraphics[width=0.8\textwidth]{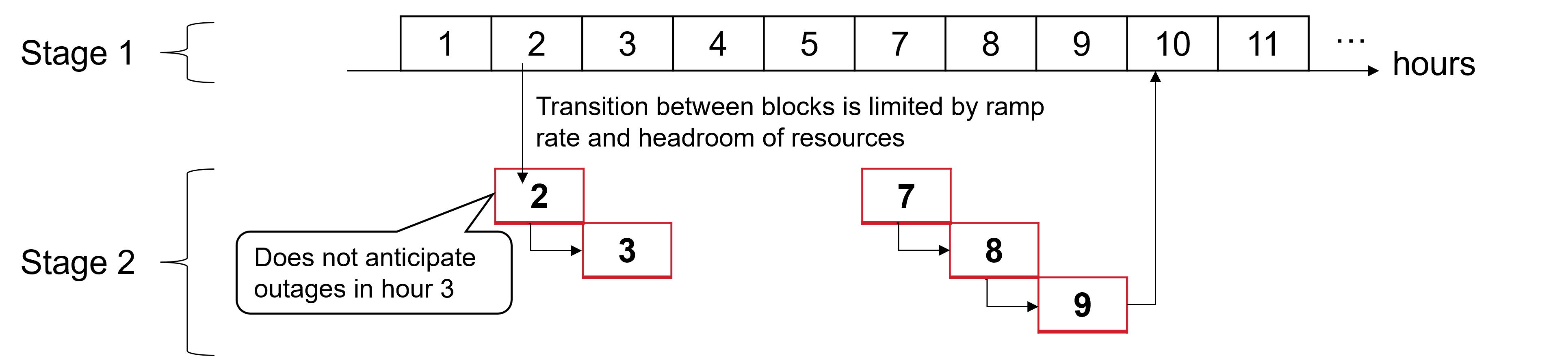}
    \caption{Overview of the Monte Carlo simulation in the reliability assessment model. The black box represents the steady-state system operation for each hour, while the red box illustrates the post-contingency system re-dispatch phase}
    \label{fig:RA_overview}
\end{figure*}

Inputs to the reliability assessment model include the generation portfolio, generator and storage characteristics, transmission topology, and time series data for load, temperature, and variable renewable energy availability. The model first generates outage scenarios for each generating asset in the system using a two-state Markov model. This model captures the dynamic evolution of individual unit-level outage states over time. During each distinct time period, the availability state of each generating asset is stochastically determined, taking into account the unit's state transition probabilities. In this study, we utilized weather-dependent state transition probabilities that are calculated using the methodology outlined in \cite{murphy_resource_2020}. After generating outage scenarios, the model conducts a Monte Carlo simulation with a two-stage structure, as shown in Fig \ref{fig:RA_overview}. In the first stage, the model solves an economic dispatch problem for each operating day to obtain steady-state dispatch setpoints for all the resources. The model does not dispatch the SATOA resources in the first stage because, by definition, these resources should maintain their full state-of-charge (SOC) levels to be used when the system has contingencies. In the second stage, for each outage scenario, the model solves a post-contingency system re-dispatch problem to assess the system’s ability to handle generator outages and determine expected unserved energy. The transition between the steady-state dispatch schedule and the post-contingency redispatch is constrained by both available headroom and the ramp rate of individual assets. In addition, the second stage simulates the post-contingency cases in a myopic way, meaning that it performs hour-by-hour simulations without anticipating future outages in the redispatch algorithm. The outcomes from all trials are then used to compute reliability metrics, such as EUE and LOLH. In this paper, we present the second stage model formulation in (9)-(28). The second stage model solves the following optimization problem for each time period during a post-contingency event.

\vspace{-0.5em}
\small
\begin{align}
    \min&\quad\sum_{it}{\left({MC}_i\left(d_{it}^{up}+d_{it}^{dn}\right)+\sum_{nt}{\left(VOLL\cdot{ens}_{nt}\right)}\right)}\label{ALEAF:eq1}
\end{align}
\normalsize
\vspace{-0.5em}
\small
\begin{align}
    &-\left(\sum_{k\left(n,\right)} f_{kt}-\sum_{k\left(,n\right)} f_{kt}\right)+\quad\sum_{\mathclap{i,Bus\left(i\right)=n}} g_{it}\quad-\qquad\sum_{\mathclap{i,Bus\left(i\right)=n,i\in\mathrm{\Psi}_{ES}}}{chg}_{it}\nonumber\\
    &\specialcell{\hfill+{ens}_{nt}-D_{nt}=0, \qquad\forall n\in N}\label{ALEAF:eq2}
\end{align}
\normalsize

\vspace{-1.5em}
\small
\begin{align}
f_{kt}&=\sum_{n}{PTDF_{kn}\cdot p_{nt}^{inj}}, 	&\forall k\in K\label{ALEAF:eq3}\\
p_{nt}^{inj}&=\sum_{\mathclap{i,Bus\left(i\right)=n}} g_{it}\,-\quad\sum_{\mathclap{i,Bus\left(i\right)=n,i\in\mathrm{\Psi}_{ES}}}{\left({chg}_{it}\,+{ens}_{nt}-D_{nt}\right)}, 	&\forall n\in N\label{ALEAF:eq4}\\
-F_k^+&\le f_{kt}\le F_k^+, 	&\forall k\in K\label{ALEAF:eq5}
\end{align}
\normalsize

\vspace{-1.5em}
\small
\begin{align}
&g_{it}\le S_{it}{CAP}_i,	&\forall i\in I\in\mathrm{\Psi}_{VRE}^{FIX}\label{ALEAF:eq6}\\
&0\le{soc}_{it}\le SOC_i^+,	&\forall i\in I\in\mathrm{\Psi}_{ES}\label{ALEAF:eq7}\\
&{soc}_{it}\!-\!{\bar{SOC}}_{i,t-1}\!-\!\epsilon_i^C{chg}_{it}\!+\!\frac{1}{\epsilon_i^D}g_{it}\!=0,	&\forall i\in I\in\mathrm{\Psi}_{ES}\label{ALEAF:eq8}
\end{align}
\normalsize
\vspace{-1.5em}
\small
\begin{align}
&g_{it}={\bar{G}}_{it}+d_{it}^{up}-d_{it}^{dn}	&\forall i\in I\label{ALEAF:eq9}\\
&\sum_{i}{C_{cit}(g_{it}-chg_{it})}-\sum_{i}{C_{cit}\left({\bar{G}}_{it}-{\bar{CHG}}_{it}\right)}\nonumber\\
&\specialcell{\hfill\le\sum_{i}{({1-C}_{cit})}({\bar{G}}_{it}-{\bar{CHG}}_{it})}\label{ALEAF:eq10}\\
&g_{it}\le C_{cit}{CAP}_i	&\forall i\in I\label{ALEAF:eq11}\\
&{chg}_{it}\le C_{cit}{CAP}_i	&\forall i\in I\label{ALEAF:eq12}\\
&d_{it}^{up}\le{CAP}_iR_i^{10}	&\forall i\in I\label{ALEAF:eq13}\\
&d_{it}^{dn}\le{CAP}_iR_i^{10}	&\forall i\in I\label{ALEAF:eq14}\\
&{chg}_{it}-{\bar{CHG}}_{it}\le{CAP}_iR_i^{10}	&\forall i\in I\label{ALEAF:eq15}\\
&{-chg}_{it}+{\bar{CHG}}_{it}\le{CAP}_iR_i^{10}	&\forall i\in I\label{ALEAF:eq16}\\
&g_{it}-{\bar{G}}_{it-1}\le{CAP}_iR_i^H	&\forall i\in I\label{ALEAF:eq17}\\
&{\bar{G}}_{it-1}-g_{it}\le{CAP}_iR_i^H	&\forall i\in I\label{ALEAF:eq18}\\
&{chg}_{it}-{\bar{CHG}}_{it-1}\le{CAP}_iR_i^H	&\forall i\in I\label{ALEAF:eq19}\\
&{\bar{CHG}}_{it-1}-{chg}_{it}\le{CAP}_iR_i^H	&\forall i\in I\label{ALEAF:eq20}
\end{align}
\normalsize

The objective function \eqref{ALEAF:eq1} minimizes the cost of load shedding and the redispatch of generators. System operating constraints are enforced through the system balance equation \eqref{ALEAF:eq2} and PTDF-based power flow constraints (\ref{ALEAF:eq3}-\ref{ALEAF:eq5}). The scheduling of variable renewable energy assets is bounded by the hourly generation profiles in \eqref{ALEAF:eq6}. The model tracks the SOC of energy storage assets with upper and lower bounds in (\ref{ALEAF:eq7}-\ref{ALEAF:eq8}). The generation redispatch in each hour of a post-contingency event is constrained by equations (\ref{ALEAF:eq9}-\ref{ALEAF:eq20}). Specifically, generation redispatch is represented as deviations from the steady-state dispatch setpoint in \eqref{ALEAF:eq9}. Constraint \eqref{ALEAF:eq10} ensures that the total amount of generation redispatch is less than the total generation lost, considering the potential of load shedding. Asset availability during the post-contingency is captured in \eqref{ALEAF:eq11} and \eqref{ALEAF:eq12} by the generator availability status, $C_{cit}$, which has a value of one if an asset is available. The deviations from the steady-state setpoints are restricted by ramp rate constraints in (\ref{ALEAF:eq15}-\ref{ALEAF:eq16}). Lastly, the inter-temporal constraints between the current and previous periods are enforced by equations (\ref{ALEAF:eq17}-\ref{ALEAF:eq20}).

\section{Case Study}\label{sec:casestudy}
This section presents a case study demonstrating the framework on an aggregated version of the Texas power system.
\subsection{Description}
\begin{figure}[!tbp]
	\centering
	\includegraphics[width=0.35\textwidth]{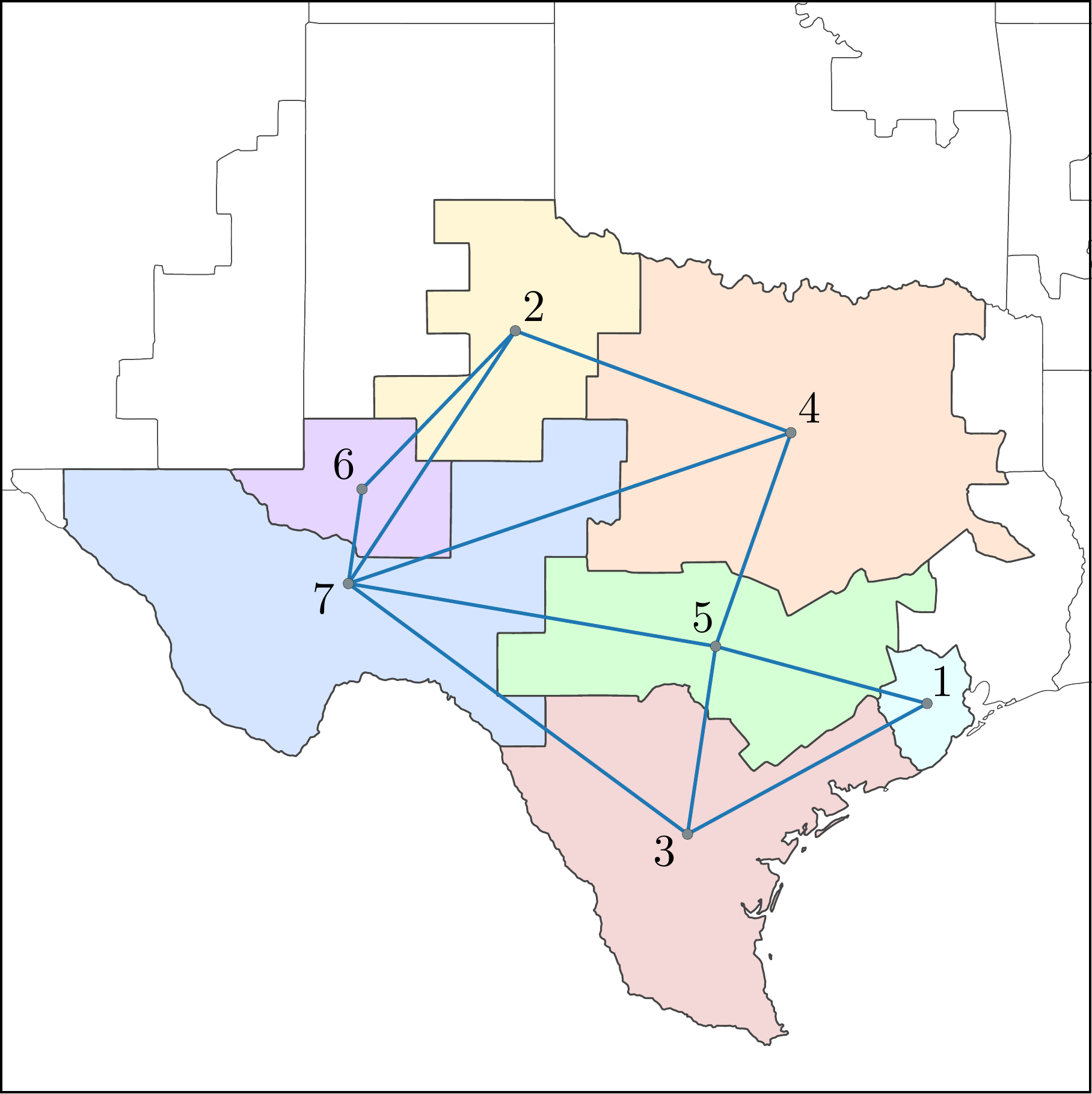}
	\caption{The network used in the case study, an aggregated ERCOT system which preserves individual generator units.}
	\label{fig:network}
\end{figure}

The network used in this case study is a simplified representation of the 
Electric Reliability Council of Texas (ERCOT) system, aggregated to seven zones as shown in Fig. \ref{fig:network}. These zones are balancing areas defined in \cite{ho_regional_2021}, from which we source our zonal load and wind and solar resource data. Table \ref{tab:cap_load} gives the non-coincident peak values for each zone.

Generation units and parameters are also taken from \cite{ho_regional_2021}, which is based on published data for real units. Each generation unit in the system is assigned to a zone but still modeled individually in both the simulation of outages and the optimization of  dispatch. Table \ref{tab:cap_load} gives the total capacity by zone and by resource type. The type ``NG'' includes natural gas simple cycle combustion turbine and combined cycle plants, and ``OGS'' represents any oil or gas steam plants.

\begin{table*}[t]
\centering
\caption{Generation capacity and peak load by zone.}
\label{tab:cap_load}
\begin{tabular}{@{}crrrrrrrrrr@{}}
\toprule
\multirow{2}{*}{Zone}&\multicolumn{9}{c}{Capacity by Resource Type (MW)}&\multirow{2}{*}{Peak Load (MW)}        \\ \cmidrule(l){2-10} 
 & Coal  & Hydro & NG & Nuclear & OGS & Solar & Storage & Wind & Total \\ \midrule
1	&   2,514	&   0	&   8,956	&   0	    &   1,163	&   480	    &   110	&   0	    &   13,223   &   19,381\\
2	&   0	    &   0	&   1,317	&   0	    &   0	    &   2,311	&   184	&   9,710	&   13,523   &   1,179\\
3	&   3,231	&   37	&   7,734	&   2,560	&   1,739	&   521	    &   311	&   8,413	&   24,546   &   12,897\\
4	&   7,047	&   124	&   15,060	&   2,400	&   5,298	&   3,158	&   150	&   3,168	&   36,405   &   30,673\\
5	&   1,615	&   315	&   6,386	&   0	    &   1,443	&   197	    &   11	&   0	    &   9,967    &   8,547\\
6	&   0	    &   0	&   2,109	&   0	    &   0	    &   2,478	&   86	&   2,563	&   7,236    &   1,840\\
7	&   0	    &   58	&   0	    &   0	    &   0	    &   3,036	&   4	&   5,927	&   9,025    &   1,838\\\midrule
Total & 14,407	&   534 &   41,562   &   4,960	&   9,642	&   12,181	&   856	&   29,782	&   113,923  &   -
\\ \bottomrule
\end{tabular}
\end{table*}

The transmission network used in this case study is an aggregated, synthetic network that is based on the existing transmission lines in the ERCOT system. The network is constructed by estimation of topology and line parameters using publicly available geographic data from the Homeland Infrastructure Foundation-Level Data \cite{us_department_of_homeland_security_homeland_2023} and aggregated to a zonal DC power flow model using the methods introduced in \cite{shi_novel_2015}. Each inter-zonal link is represented by its reactance and thermal capacity, as shown in Table \ref{tab:branches}. 

\begin{table}[!tbp]
\centering
\caption{Aggregated transmission link parameters.}
\label{tab:branches}
\begin{tabulary}{0.7\linewidth}{@{}LLRR@{}}
\toprule
From Zone	& To Zone	  & Reactance (p.u.)	& Thermal Capacity (MW)\\ \midrule
1	&   3	&   0.0042    &   5,294     \\
1	&   5	&   0.0024    &   5,109     \\
2	&   4	&   0.0018    &   14,463     \\
2	&   6	&   0.0036    &   6,610     \\
2	&   7	&   0.0016    &   3,497     \\
3	&   5	&   0.0010    &   10,478     \\
3	&   7	&   0.0115    &   283     \\
4	&   5	&   0.0015    &   9,731     \\
4	&   7	&   0.0177    &   3,384     \\
5	&   7	&   0.0126    &   3,164     \\
6	&   7	&   0.0033    &   2,928 \\ \bottomrule
\end{tabulary}
\end{table}

\subsection{Transmission and Storage Expansion}
The JHSMINE transmission and storage expansion model is applied to this system to find optimal expansion decisions for each of three cases: transmission expansion (TEP), battery storage expansion (BEP), and joint  battery storage and transmission expansion (BTEP). The single peak load hour of each day is considered, for a total of 365 unique operating conditions. Due to this daily resolution, generator ramp constraints and battery charging and discharging ramp constraints are not modeled in the expansion problem, but they are considered in the detailed reliability assessment. The TEP problem represents a TPP scenario in which the interface between Zone 3 and 7 has been identified as a limiting constraint, and planners seek the optimal sizing of the upgrade based on a least-cost expansion formulation. This link may be upgraded to increase its thermal capacity at a cost of 1.6 thousand USD per MW-mile, with the MW increase modeled as a continuous variable between 0 and 3,000 MW. The BEP problem represents an alternative approach which considers energy storage as candidate upgrades throughout the system. We consider a generic four-hour duration battery energy storage system (BESS) that can be installed in any zone in 400 MW increments up to 1,200 MW total, with an investment cost of 0.35 million USD per MWh. The BTEP problem represents a combination of the TEP and BEP, allowing co-optimization of link 3-7 upgrade and storage investments.

Table \ref{tab:jhsmine_results} shows the optimal expansion results of each problem. The TEP resulted in an upgrade to increase the capacity of this link by 1,495 MW, at a cost of 80.46 million USD, resulting in very large operational cost savings. The reason for this is that the alleviation of this severe transmission constraint allows the renewable production in Zone 7 to be sent to other zones, significantly reducing curtailment of these resources. No storage solution in this example was able to offer this benefit, or any operational benefit outweighing the investment cost, as the results of the BEP show no storage investments. The BTEP result further shows that no storage investment resulted in a net benefit, even if coupled with the transmission upgrade. It is important to note that no load shedding is observed throughout the 365 periods of operation in the expansion analysis with JHSMINE.

\begin{table}[!tbp]
\centering
\caption{Expansion planning results.}
\label{tab:jhsmine_results}
\begin{tabular}{p{3.5cm}p{1.2cm}p{1.2cm}p{1.2cm}}
\toprule
Case    &   BTEP    &   BEP &   TEP \\ \midrule
Annual operation cost (B\$) &   4.337 &   6.657 &   4.337   \\ \midrule
Line investment cost (M\$) &    80.46 & 0   & 80.46   \\ \midrule
Storage investment cost (M\$) &    0 & 0   & ---   \\ \midrule
Selected lines  &     3-7 (1,495 MW) & 0   &  3-7 (1,495 MW)   \\ \midrule
Selected storage &    --- & ---   & ---  
\\ \bottomrule
\end{tabular}
\end{table}

\subsection{Assessing Reliability Benefits of Potential Expansion Decisions}
Next, the reliability is assessed using A-LEAF. To better understand the reliability contributions associated with SATOA additions, ten candidate storage installations are examined in Zone 3, including two- and four-hour systems with capacities ranging from 300 MW to 1500 MW. To assess the reliability, we first generate a set of 1,000 random generator outage scenarios for each hour of the year, then we simulate operations under each of the configurations for each outage scenario. Finally, for each configuration, we compute the EUE and LOLH from the outcomes of all scenarios. The reliability assessment model simulates system dispatch across all 8,760 hourly periods in the year. The inclusion of these operational details and contingency modeling enables the reliability outcomes under each configuration to be captured with greater accuracy.

\begin{figure}[!tbp]
    \centering
    \includegraphics[width=0.5\textwidth]{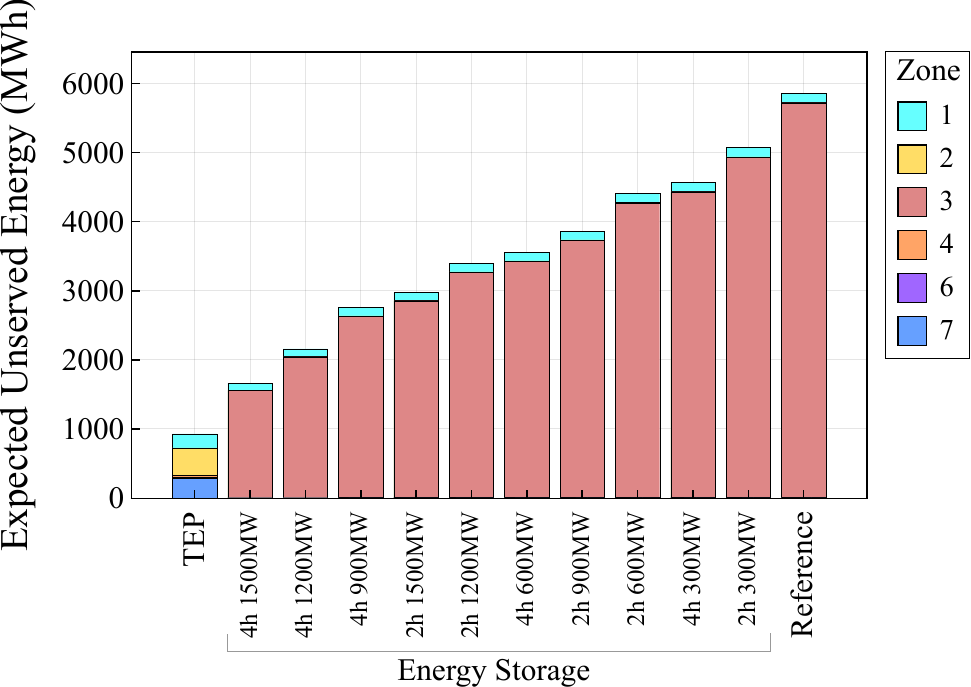}
    \caption{Annual expected unserved energy (EUE) in each zone under the Transmission Expansion (TEP), Battery Expansion (BEP, by zone in which the battery is installed), and Reference cases.}
    \label{fig:eue_annual}
\end{figure}

In contrast to the expansion model, unserved load is observed under each configuration. Fig. \ref{fig:eue_annual} shows the EUE by zone for each configuration. The TEP result gives a significant improvement compared to the reference case. The energy storage investments demonstrate improvements which increase with storage capacity, from a slight improvement over the reference case with a 2 hour, 300 MW system, to a large improvement with a 4 hour, 1500 MW system. The storage investments do not outperform the transmission expansion planning results, however. Table \ref{tab:eue} shows the EUE values for the reference, TEP, and three of the storage configurations in greater detail. The values in the table are in MWh, followed by the relative change compared to the reference case as a percentage in parentheses. In the reference and storage cases, the EUE in Zone 3 is the largest of any zone. The TEP result completely eliminates the unserved energy in Zone 3, though several other zones experience increases in EUE. This demonstrates that transmission upgrades may lead to significant changes in the patterns of generation dispatch and power flow through a network, potentially influencing the distribution of reliability throughout the network.

\begin{table*}[!tbp]
\centering
\caption{Annual expected unserved energy by zone under the reference, transmission expansion, and three storage cases.}\label{tab:eue}
\begin{tabular}{@{}rrrrrr@{}}
\toprule
\multicolumn{1}{c}{}                       & \multicolumn{5}{c}{Annual Expected Unserved Energy (MWh)}     \\ \cmidrule(l){2-6} 
\multicolumn{1}{c}{\multirow{-2}{*}{Zone}} & Reference  & TEP (\emph{Relative})  & \stackanchor{Storage}{(4h 1500MW)} (\emph{Relative}) & \stackanchor{Storage}{(4h 600MW)} (\emph{Relative}) & \stackanchor{Storage}{(2h 600MW)} (\emph{Relative}) \\ \midrule
1                                          & 138.4  & 201.9 \hbox to 1.25cm{\hfil(+45.8\%) }  & 100.3	\hbox to 1.25cm{\hfil(-27.6\%)} & 125.9  \hbox to 1.25cm{\hfil(-9.0\%)}  & 130.8     \hbox to 1.25cm{\hfil(-5.5\%)}\\
2                                          & 5.5	& 399.9 \hbox to 1.25cm{\hfil(+7197.5\%)} & 6.9	    \hbox to 1.25cm{\hfil(+25.1\%)} & 6.5	 \hbox to 1.25cm{\hfil(+19.1\%)} &  5.7	     \hbox to 1.25cm{\hfil(+3.5\%)}\\
3                                          & 5717.1 & 0.0	\hbox to 1.25cm{\hfil(-100.0\%)}  & 1552.6  \hbox to 1.25cm{\hfil(-72.8\%)} & 3423.0 \hbox to 1.25cm{\hfil(-40.1\%)} &  4267.3   \hbox to 1.25cm{\hfil(-25.4\%)}\\
4                                          & 0.0	& 27.3  \hbox to 1.25cm{\hfil---}         & 0.0	    \hbox to 1.25cm{\hfil---}       & 0.0	 \hbox to 1.25cm{\hfil---}       &  0.0	     \hbox to 1.25cm{\hfil---}      \\
5                                          & 0.0	& 0.0	\hbox to 1.25cm{\hfil---}         & 0.0	    \hbox to 1.25cm{\hfil---}       & 0.0	 \hbox to 1.25cm{\hfil---}       &  0.0	     \hbox to 1.25cm{\hfil---}      \\
6                                          & 0.0	& 14.2  \hbox to 1.25cm{\hfil---}         & 0.0	    \hbox to 1.25cm{\hfil---}       & 0.0	 \hbox to 1.25cm{\hfil---}       &  0.0	     \hbox to 1.25cm{\hfil---}      \\
7                                          & 3.4	& 282.9 \hbox to 1.25cm{\hfil(+8264.5\%)} & 1.7	    \hbox to 1.25cm{\hfil(-49.3\%)} & 2.4	 \hbox to 1.25cm{\hfil(-28.3\%)} &  3.0	     \hbox to 1.25cm{\hfil(-11.9\%)} \\\midrule
Systemwide                                 & 5864.4 & 926.2 \hbox to 1.25cm{\hfil(-84.2\%)}  & 1661.4  \hbox to 1.25cm{\hfil(-71.7\%)} & 3557.9 \hbox to 1.25cm{\hfil(-39.3\%)} &  4406.7   \hbox to 1.25cm{\hfil(-24.9\%)}\\\bottomrule
\end{tabular}
\end{table*}

\begin{figure}[!tbp]
    \centering
    \includegraphics[width=0.5\textwidth]{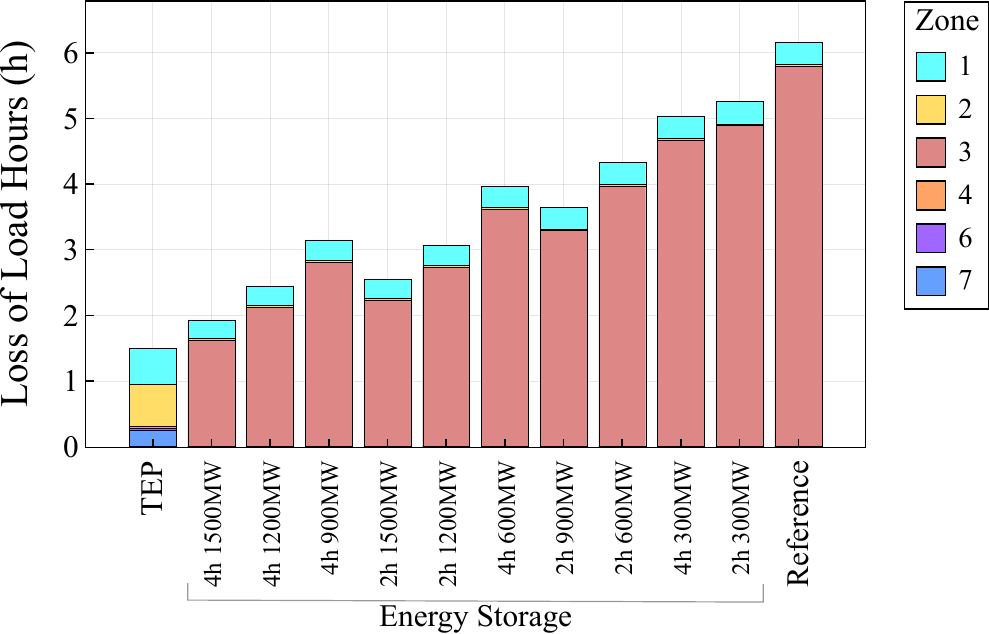}
    \caption{Annual loss of load hours (LOLH) in each zone under the Transmission Expansion (TEP), Battery Expansion (BEP, by zone in which the battery is installed), and Reference cases.}
    \label{fig:lolh_annual}
\end{figure}

Similarly, Fig. \ref{fig:lolh_annual} shows the LOLH by zone for each configuration, and  Table \ref{tab:lolh} gives additional detail. The TEP again gives the best performance under this metric. The storage solutions result in relative improvements in LOLH which are quite similar to their improvements in EUE. The TEP solution also results in similar relative improvements in LOLH as it does in EUE. 

\begin{table*}[!tbp]
\centering
\caption{Annual loss of load hours by zone under the reference, transmission expansion, and three storage cases.}\label{tab:lolh}
\begin{tabular}{@{}rrrrrr@{}}
\toprule
\multicolumn{1}{c}{}                       & \multicolumn{5}{c}{Annual Loss of Load Hours (h)}     \\ \cmidrule(l){2-6} 
\multicolumn{1}{c}{\multirow{-2}{*}{Zone}} & Reference  & TEP (\emph{Relative})  & \stackanchor{Storage}{(4h 1500MW)} (\emph{Relative}) & \stackanchor{Storage}{(4h 600MW)} (\emph{Relative}) & \stackanchor{Storage}{(2h 600MW)} (\emph{Relative}) \\ \midrule
1                                          & 0.34 & 0.54 \hbox to 1.25cm{\hfil(+57.4\%) }  & 0.27 \hbox to 1.25cm{\hfil(-21.3\%)} & 0.32 \hbox to 1.25cm{\hfil(-6.7\%)}  & 0.33 \hbox to 1.25cm{\hfil(-3.5\%)}\\
2                                          & 0.03 & 0.65 \hbox to 1.25cm{\hfil(+2492.0\%)} & 0.03 \hbox to 1.25cm{\hfil(+8.0\%)}  & 0.03 \hbox to 1.25cm{\hfil(+20.0\%)} & 0.03 \hbox to 1.25cm{\hfil(+12.0\%)}\\
3                                          & 5.79 & 0.00 \hbox to 1.25cm{\hfil(-100.0\%)}  & 1.62 \hbox to 1.25cm{\hfil(-71.9\%)} & 3.61 \hbox to 1.25cm{\hfil(-37.6\%)} & 3.96 \hbox to 1.25cm{\hfil(-31.5\%)}\\
4                                          & 0.00 & 0.03 \hbox to 1.25cm{\hfil---}         & 0.00 \hbox to 1.25cm{\hfil---}       & 0.00 \hbox to 1.25cm{\hfil---}       & 0.00 \hbox to 1.25cm{\hfil---}      \\
5                                          & 0.00 & 0.00 \hbox to 1.25cm{\hfil---}         & 0.00 \hbox to 1.25cm{\hfil---}       & 0.00 \hbox to 1.25cm{\hfil---}       & 0.00 \hbox to 1.25cm{\hfil---}      \\
6                                          & 0.00 & 0.03 \hbox to 1.25cm{\hfil---}         & 0.00 \hbox to 1.25cm{\hfil---}       & 0.00 \hbox to 1.25cm{\hfil---}       & 0.00 \hbox to 1.25cm{\hfil---}      \\
7                                          & 0.01 & 0.25 \hbox to 1.25cm{\hfil(2644.4\%)}  & 0.00 \hbox to 1.25cm{\hfil(-55.6\%)} & 0.01 \hbox to 1.25cm{\hfil(-44.4\%)} & 0.01 \hbox to 1.25cm{\hfil(0.0\%)} \\\midrule
Systemwide                                 & 6.16 & 1.49 \hbox to 1.25cm{\hfil(-75.8\%)}   & 1.93 \hbox to 1.25cm{\hfil(-68.8\%)} & 3.97 \hbox to 1.25cm{\hfil(-35.7\%)} & 4.33 \hbox to 1.25cm{\hfil(-29.7\%)} \\\bottomrule
\end{tabular}
\end{table*}

The lost load events which make up the annual metrics in Tables \ref{tab:eue} and \ref{tab:lolh} occur on a relatively small number of days. This can be seen in Fig. \ref{fig:eue_sys}, which shows the EUE by zone in the TEP, BEP, and Reference cases on the four days with the highest lost load.  August 2nd and 1st experience the most severe shortfalls, accounting for 53\% of the yearly EUE in the reference case. The behavior exhibits a similar pattern across the four days.

\begin{figure*}[!tbp]
    \centering
    \includegraphics[width=\textwidth]{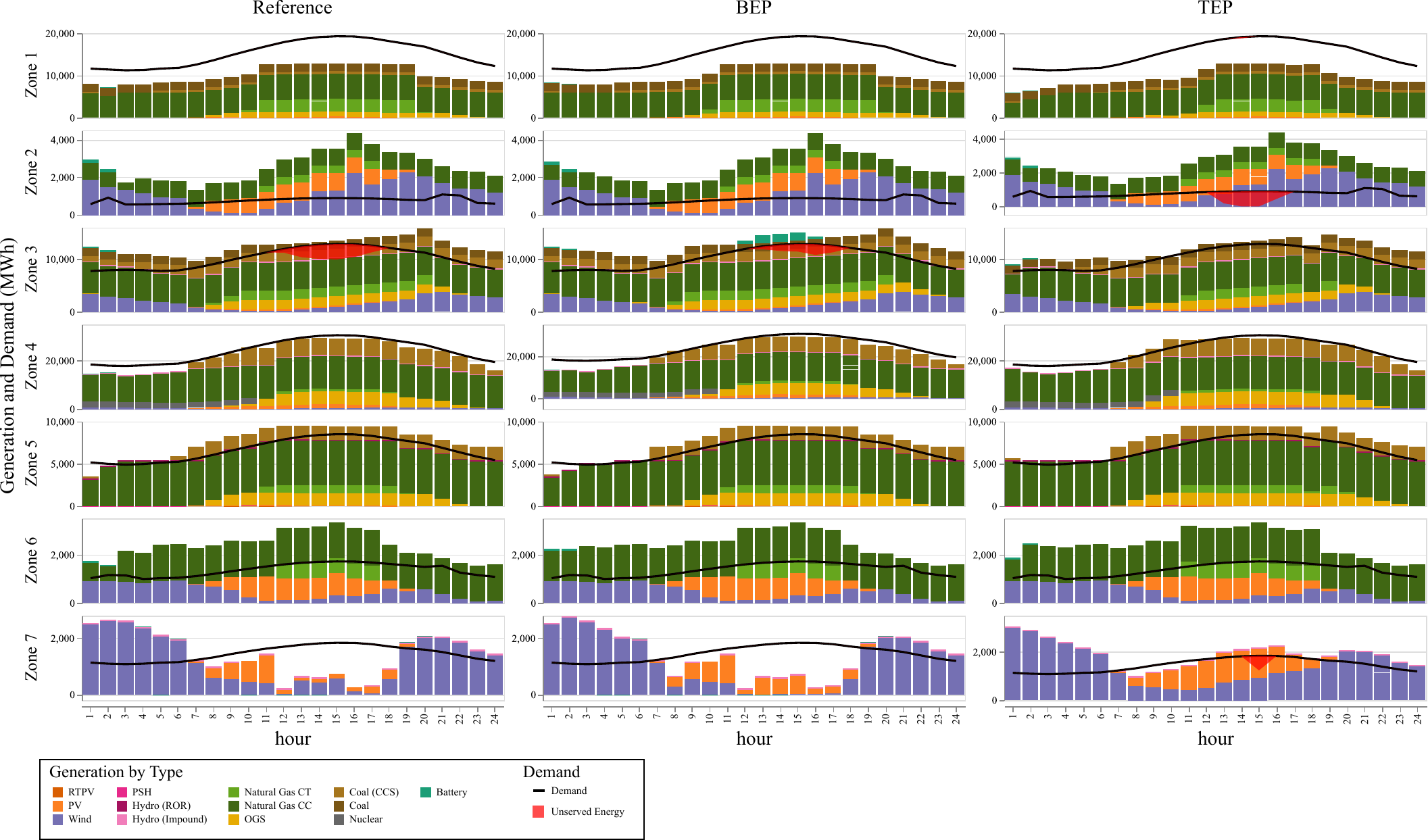}
    \caption{Example of a single sample outage scenario on August 2. Each plot shows the generation, demand, and unserved energy. The rows represent each zone, and the columns represent the Reference, storage (4 h 1500 MW), and TEP configurations.}
    \label{fig:dispatch214}
\end{figure*}

\begin{figure}[!tbp]
    \centering
    \includegraphics[width=0.4\textwidth]{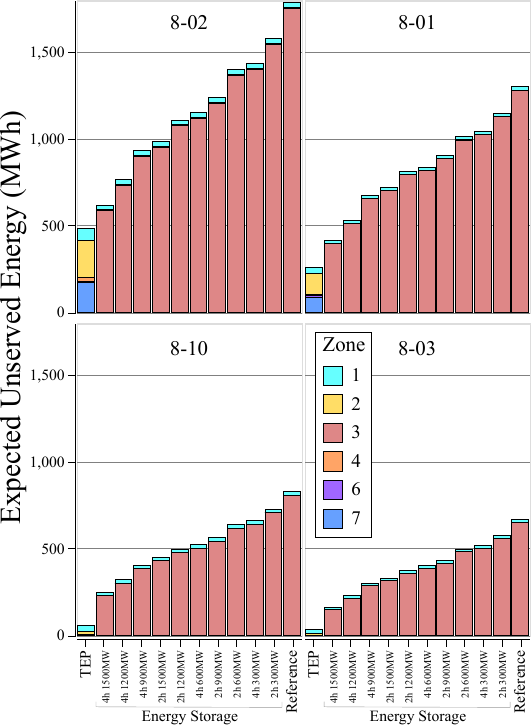}
    \caption{Expected unserved energy (EUE) in each zone by day, for the four days which have the highest EUE under the Transmission Expansion (TEP), Storage in Zone 3, and Reference cases.}
    \label{fig:eue_sys}
\end{figure}

Fig. \ref{fig:dispatch214} shows the hourly dispatch by generation type for one sample outage scenario on August 2nd. The rows of the figure make up the results for all the zones, and the columns represent the results under the Reference, BEP in Zone 3, and TEP configurations, respectively. The realization of the stochastic characteristics in this single sample are identical between the three configurations; any differences across the three columns are only the result of the storage or transmission additions. Overlaid on each plot are the hourly demand profiles in black, along with a red area representing any unserved energy in the zone. The system outcomes shown in Fig. \ref{fig:dispatch214}, though only representing a single sample on a single day (Aug. 2nd) are illustrative of several observed across reliability metrics. In the storage configuration, the new storage installed in Zone 3 is discharged in hours 12 through 16, contributing to a decrease in unserved energy in this zone. The reduced curtailment of wind and photovoltaic generation in Zone 7 can be seen in the TEP solution, as well as the unserved load event that occurs in hour 15 in that zone.

\subsection{Cost and Reliability}
\begin{figure}[!tbp]
    \centering
    \includegraphics[width=0.45\textwidth]{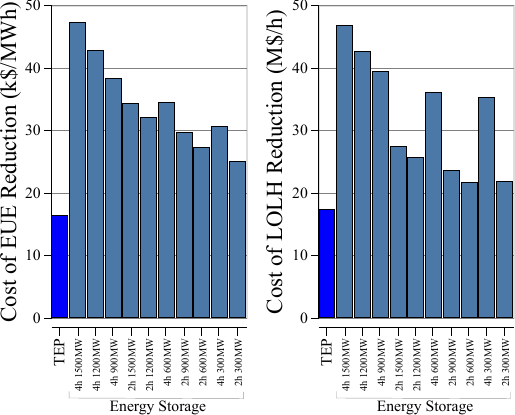}
    \caption{Capital cost of reduction of expected unserved energy and loss of load hours under the BEP configurations with a battery in Zones 3 and 5 and the TEP configuration.}
    \label{fig:reliability_cost}
\end{figure}

In order to compare the reliability outcomes between the expansion configurations, we compute the cost of reduction of both expected unserved energy and loss of load hours by dividing the annual investment costs shown in \ref{tab:upgrades} by the reduction in each metric compared with the reference case. Fig. \ref{fig:reliability_cost} shows these costs for the storage configurations in Zone 3 and for the TEP configuration. The TEP configuration offers the lowest cost of reliability improvements in both EUE and LOLH. The costs of reliability improvements under the storage options are slightly higher in this case. The smaller storage units, however, give more comparable costs of improvement and offer lower capital costs, potentially allowing more to be placed throughout the network in situations where more distributed application is advantageous. Note that this comparison does not include other values, such as operational cost savings, that could result from these investments. Operational savings may often present an advantage of transmission investments compared to SATOA investments, since SATOA must reserve its capacity for essential transmission services.

\begin{table}[!tbp]
    \centering
    \caption{Annualized cost of each upgrade.}\label{tab:upgrades}
    \begin{tabular}{@{}lr@{}}
    \toprule
    Upgrade & Cost (M\$/annum)\\ \midrule
    Storage: 2h, 300MW	& 19.8 \\
    Storage: 2h, 600MW	& 39.6 \\
    Storage: 2h, 900MW	& 59.5 \\
    Storage: 2h, 1,200MW	& 79.3 \\
    Storage: 2h, 1,500MW	& 99.1 \\
    Storage: 4h, 300MW	& 39.6 \\
    Storage: 4h, 600MW	& 79.3 \\
    Storage: 4h, 900MW	& 118.9 \\
    Storage: 4h, 1,200MW	& 158.6 \\
    Storage: 4h, 1,500MW	& 198.2 \\
    TEP	& 80.7 \\\bottomrule
\end{tabular}
\end{table}

\section{Conclusions}\label{sec:conclusions}
This paper presents a two-stage analysis framework to quantify the reliability benefits of transmission assets, including SATOA. The framework consists of an optimal storage and transmission expansion planning methodology formulated as a mixed integer linear program using a DC power flow representation, along with a probabilistic reliability assessment methodology that simulates system operations under stochastic generator outage scenarios.

This framework is used to demonstrate that transmission and storage each may be identified as appropriate solutions for different aspects of reliability needs. Traditional least cost expansion tools provide valuable insights but may not capture factors which influence reliability, such as correlated generator outages, and they need to be coupled or integrated with detailed reliability assessment to quantitatively evaluate of the reliability benefits of storage as a transmission asset. The case study presented here demonstrated that the upgrade of a severely constrained transmission corridor connecting a zone with excess renewable generation to a high-load zone with poor reliability was optimal from a least-cost expansion perspective and also provided significant reliability benefits. Storage as a transmission asset was not able to outperform the reliability contributions of the transmission upgrade in this scenario, but large storage installations achieved nearly comparable improvements of 71.7\% in EUE and 68.8\% in LOLH, compared to the TEP solution's 84.2\% and 75.8\%, respectively. In addition, smaller storage additions showed capital costs per unit improvement of 25.1 thousand USD/MWh EUE reduction and 21.9 million USD/h LOLH reduction, compared to 16.3 thousand USD/MWh and 17.3 million USD/h, respectively, under the TEP solution. Given the lower total cost and potentially greater feasibility of installation, storage solutions for incremental improvements may offer advantages in certain cases. This underscores the need for planning tools which accurately assess both the economic and reliability contributions of both traditional transmission and storage as transmission only solutions for quantitative comparisons of which solution best meets the planning needs at hand.

This work also points to several important future directions. The inability of SATOA to bring operational savings due to its exclusion from market participation may be a barrier to its adoption. The reliability implications of allowing partial market participation should be studied. In the case study, the aggregated network obscures detailed power flow constraints and therefore potentially underestimates transmission congestion and its reliability impacts. A finer network resolution would more accurately capture whether the reliability benefits of storage are primarily realized locally or alternatively whether they can be allocated across the system via existing transmission infrastructure. Incorporating a forced outage model for transmission lines would also improve the characterization of the reliability benefits of each transmission asset. Future work includes the application of the framework with more detailed transmission network models to reflect these complex effects of congestion as well as the incorporation of models of weather-dependent transmission outage behavior. Additional future work will examine the participation of energy storage both as a transmission and market asset, identifying operational constraints which can enable this dual use while maintaining reliability benefits.

\section*{Nomenclature}
\addcontentsline{toc}{section}{Nomenclature}
\footnotesize
\subsection*{JHSMINE Sets and Indices}
\begin{IEEEdescription}[\IEEEusemathlabelsep\IEEEsetlabelwidth{$G$}]
\item[$F$] Fuel types, index $f$.
\item[$G$] Generation technologies, index $g$.
\item[$H$]	Hours within a representative period (e.g., 24 hours), index $h$.
\item[$I$]	Buses, index $i$.
\item[$J$]	Energy storages, index $j$.
\item[$K$]	Generators, index $k$.
\item[$L$]	Transmission lines, index $l$.
\item[$RP$]	Representative periods, index $rp$.
\item[$S$]	Scenarios, index $s$.
\item[$T$]	Hours of a planning period (e.g., 8760 hours), index $t$.
\item[$W$]	States/Provinces, index $w$.
\item[$Y$]	Years, index $y$.
\end{IEEEdescription}
\subsection*{JHSMINE Parameters}
\begin{IEEEdescription}[\IEEEusemathlabelsep\IEEEsetlabelwidth{$CTAXs,y,w$}]
\item[$CTAX_{s,y,w}$] Carbon price or tax of state $w$ in the scenario tree node $(s,y)$, unit \$/ton CO\textsubscript{2}e.
\item[$D_y$]	            Discounting factor of year $y$ to the beginning of the planning horizon.
\item[$DA_y$]	            Accumulative discounting factor of year $y$ to the beginning of the planning horizon. 
\item[$EER_j$] 	        Emission rate of the storage facility $j$, unit ton CO\textsubscript{2}e/MWh.
\item[$EEXC_{s,y,j}$] 	    Expansion cost of the storage facility $j$ in scenario tree node $(s,y)$, unit M\$. 
\item[$EFOM_j$]	        Fixed O\&M cost of the storage $j$, unit \$/MW-year.
\item[$EHR_j$] 	        Heat rate of the storage $j$, unit MMBTU/MWh.
\item[$ESAL_{s,y,j,s^\prime,y^\prime}$] 	Salvage revenue of storage facility $j$ if expanded in the scenario tree node $(s^\prime,y^\prime)$ and retired in node $(s,y)$, unit M\$.
\item[$EVOM_j$] 	        Variable O\&M cost of the storage $j$, unit \$/MWh.
\item[$FC_{s,y,rp,h,f}$]	    Price of the fuel $f$ at hour $h$ of representative period rp in scenario tree node $(s,y)$, unit \$/MMBTU.
\item[$GER_k$]	            Emission rate of the generator $k$, unit tonCO\textsubscript{2}e/MWh.
\item[$GEXC_{s,y,k}$]	    The expansion cost of the generator $k$ in scenario tree node $(s,y)$, unit M\$.
\item[$GFOM_k$] 	        Fixed operating and maintenance cost of the generator $k$, unit \$/MW-year.
\item[$GHR_k$]	            Average heat rate of generator $k$, unit MMBTU/MWh.
\item[$GNPL_k$]	        Nameplate capacity of the generator $k$, unit MW. 
\item[$GSAL_{s,y,k,s^\prime,y^\prime}$]	The salvage revenue of the generator $k$ if it is expanded in scenario tree node $(s^\prime,y^\prime)$ and retired in the node $(s,y)$, unit M\$.
\item[$GSUC_k$]	        Start-up cost of generator $k$ per unit of the capacity, unit \$/MW.
\item[$GVOM_k$]	        Variable operating and maintenance cost of the generator $k$, unit \$/MWh.
\item[$LEXC_{s,y,l}$] 	    Expansion cost of the transmission line $l$ in the scenario tree node $(s,y)$, unit M\$.
\item[$LSAL_{s,y,l,s^\prime,y^\prime}$] 	Salvage revenue of the transmission line $l$ if it is built in the scenario tree node $(s^\prime,y^\prime)$ and retired in the node $(s,y)$, unit M\$.
\item[$RACP_w$] 	        Alternative compliance penalty for RPS of state $w$, unit \$/MWh 
\item[$RPW_{y,rp}$]	        \# of periods represented by representative period $rp$ in year $y$.
\item[$SP_{s,y}$]	        Scenario probability, unitless.
\item[$VOLL$]	            Value of lost load, unit \$/MWh.
\end{IEEEdescription}
\subsection*{JHSMINE Variables}
\subsubsection*{Expansion and Retirement}
\begin{IEEEdescription}[\IEEEusemathlabelsep\IEEEsetlabelwidth{$CTAXs,y,w$}]
\item[$eincexp_{s,y,j}$] Storage incremental expansion, 1 if the storage $j$ becomes commissioned in $(s,y)$, binary, unitless.
\item[$eincret_{s,y,j}$]	Storage incremental retirement, 1 if the storage $j$ is decommissioned in $(s,y)$, binary, unitless.
\item[$gincexp_{s,y,k}$] 	Generator incremental expansion, 1 if the generator $k$ becomes commissioned in $(s,y)$, binary, unitless.
\item[$gincret_{s,y,k}$]	Generator incremental retirement, 1 if the generator $k$ is decommissioned in $(s,y)$, binary, unitless.
\item[$lincexp_{s,y,l}$] 	Transmission line incremental expansion, 1 if the transmission line $l$ becomes commissioned in $(s,y)$, binary, unitless.
\item[$lincret_{s,y,l}$]	Transmission line incremental retirement, 1 if the transmission line $l$ is decommissioned in $(s,y)$, binary, unitless.
\end{IEEEdescription}
\subsubsection*{Operations}
\begin{IEEEdescription}[\IEEEusemathlabelsep\IEEEsetlabelwidth{$CTAXs,y,w$}]
\item[$edis_{s,y,rp,h,j}$] Discharge of the storage $j$ at the hour $h$, nonnegative, unit MW.

\item[$gopt_{s,y,rp,h,k}$] 	Power output of the generator $k$ at the hour $h$, nonnegative, unit MW.
\item[$gsup_{s,y,rp,h,k}$] 	Start-up action of the generator $k$ at the beginning of the hour $h$, binary, unitless.  
\item[$nload_{s,y,rp,h,i}$] 	Load shedding at bus $i$ at the hour $h$, nonnegative, unit MW.
\item[$nrps_{s,y,rp,h,w}$] 	Non-compliance with RPS policy, unit MW, nonnegative.
\end{IEEEdescription}
\subsubsection*{Objective Function}
\begin{IEEEdescription}[\IEEEusemathlabelsep\IEEEsetlabelwidth{$fomc_{s,y}$}]
\item[$obj$] Objective function, unit \$.
\item[$invc_{s,y}$]	Investment cost occurs at the scenario tree node $(s,y)$, unit \$.
\item[$oprc_{s,y}$]	Operation cost occurs at the scenario tree node $(s,y)$, unit \$.
\item[$fomc_{s,y}$]	Fixed O\&M cost occurs at the scenario tree node $(s,y)$, unit \$.
\item[$fuelc_{s,y}$]	Fuel cost occurs at the scenario tree node $(s,y)$, unit \$.
\item[$vomc_{s,y}$]	Variable cost occurs at the scenario tree node $(s,y)$, unit \$.
\item[$stuc_{s,y}$]	Start-up cost occurs at the scenario tree node $(s,y)$, unit \$.
\item[$ctax_{s,y}$]	Carbon tax payment occurs at the scenario tree node $(s,y)$, unit \$.
\item[$voll_{s,y}$]	Lost load cost occurs at the scenario tree node $(s,y)$, unit \$.
\item[$rpsc_{s,y}$]	Cost of renewable portfolio standards non-compliance penalty, occurs at the scenario tree node $(s,y)$, unit \$.
\end{IEEEdescription}
\subsection*{A-LEAF Sets and Indices}
\begin{IEEEdescription}[\IEEEusemathlabelsep\IEEEsetlabelwidth{$\mathrm{\Psi}_{VRE}^{FIX}\in I$}]
\item[$d\in D$]	Set of representative day groups.
\item[$i\in I$]	Set of generator technologies.
\item[$h\in H$]	Set of hour periods.
\item[$t\in T$]	Set of sub-hour periods.
\item[$c\in C$]	Set of risk scenario.
\item[$k\in K$]	Set of transmission lines.
\item[$n\in N$]	Set of buses.
\item[$\mathrm{\Psi}_{ES}\in I$]	Set of energy storage generators.
\item[$\mathrm{\Psi}_{VRE}^{FIX}\in I$]	Set of variable renewable energy generators with fixed generation profiles
\end{IEEEdescription}
\subsection*{A-LEAF Parameters}
\begin{IEEEdescription}[\IEEEusemathlabelsep\IEEEsetlabelwidth{${\bar{CHG}}_{it}$}]
\item[$C_{cit}$]	Generator availability status (1: available, 0: unavailable)
\item[$CAP_i$]	Power capacity, index $i$, unit MW
\item[${\bar{CHG}}_{it}$]	Steady-state energy storage charging, unit MW
\item[$D_{nt}$]	Power demand, indices $n$,$t$, unit MW
\item[$\epsilon_i^C$]	Energy storage charging efficiency, index $i$
\item[$\epsilon_i^D$]	Energy storage discharging efficiency, index $i$
\item[$F_k^+$]	Transmission capacity, index $k$, unit MW
\item[${\bar{G}}_{it}$]	Steady-state generator dispatch setpoint, indices $i,t$, unit MW
\item[${MC}_i$]	Marginal variable cost, index $i$, unit \$
\item[$PTDF_{kn}$]	Power transfer distribution factor, indices $k$,$n$
\item[$R_i^{10}$]	10 mins ramp capacity, index $i$, unit MW
\item[$R_i^H$]	Hourly ramp capability, index $i$, unit MW
\item[$S_{it}$]	Renewable generation profile, indices $i$,$t$, unit MW
\item[$SOC_i^+$]	Maximum state-of-charge, index $i$
\item[$VOLL$]	Value of lost load, unit \$/MWh.
\end{IEEEdescription}

\subsection*{A-LEAF Variables}
\begin{IEEEdescription}[\IEEEusemathlabelsep\IEEEsetlabelwidth{$p_{nt}^{inj}$}]
\item[${chg}_{it}$]	Storage charging, indices $i$,$t$, unit MW
\item[${ens}_{nt}$]	Energy not served, indices $n$,$t$, unit MWh
\item[$f_{kt}$]	Power flow, indices $k$,$t$, unit MW
\item[$g_{it}$]	Power output, indices $i$,$t$, unit MW
\item[$p_{nt}^{inj}$]	Net power injection, indices $n$,$t$, unit MW
\item[$d_{it}^{up}$]	Generator redispatch up, indices $i$,$t$, unit MW
\item[$d_{it}^{dn}$]	Generator redispatch down, indices $i$,$t$, unit MW
\end{IEEEdescription}

\normalsize

\bibliography{bibtex/bib/IEEEabrv.bib,bibtex/bib/references.bib}{}
\bibliographystyle{IEEEtran}

\end{document}